\documentclass[12pt]{article}
\usepackage{times}
\usepackage{geometry}
\usepackage{setspace}
\usepackage{caption}
\usepackage{booktabs}
\usepackage{threeparttable}

\usepackage{amsmath}
\usepackage{mathtools}
\usepackage{amssymb} 
\usepackage{amsfonts}
\usepackage{graphicx}

\usepackage[doi=false,isbn=false,url=false,eprint=false,date=year,style=nature]{biblatex}
\usepackage{url}
\usepackage{hyperref}
\hypersetup{colorlinks=true,
            citecolor=blue,
            linkcolor=blue,
            urlcolor=blue,
            breaklinks=true
}

\usepackage{authblk}

\title{Societal citations undermine the function of the science reward system}

\author
{Xiaokai Li$^{1}$, An Zeng$^{1\ast}$, and Ying Fan$^{1\ast}$
\\

$^{1}$School of Systems Science, Beijing Normal University, Beijing 100875, China

$^{\ast}$Corresponding authors; Email: anzeng@bnu.edu.cn; yfan@bnu.edu.cn

}

\date{}

\begin{document}
\captionsetup[figure]{font={stretch=1.2},labelfont={bf},labelformat={default},labelsep=space,name={Fig.}}
\captionsetup[table]{font={stretch=1.2},labelfont={bf},labelsep=space,name={Tab.}}

\maketitle

\section*{\centering Abstract}

Citations in the scientific literature system do not simply reflect relationships between knowledge but are influenced by non-objective and societal factors.
Citation bias, irresponsible citation, and citation manipulation are widespread and have become a serious and growing problem.
However, it has been difficult to assess the consequences of mixing societal factors into the literature system because there was no observable literature system unmixed with societal factors for comparison.
In this paper, we construct a mathematical theorem network, representing a logic-based and objective knowledge system, to address this problem.
By comparing the mathematical theorem network and the scientific citation networks, we find that these two types of networks are significantly different in their structure and function.
In particular, the reward function in citation networks is impaired:
The scientific citation network fails to provide more recognition for more disruptive results, while the mathematical theorem network can achieve.
We develop a network generation model that can create two types of links---logical and societal---to account for these differences.
The model parameter $q$, which we call the human influence factor, can control the number of societal links and thus regulate the degree of mixing of societal factors in the networks.
Under this design, the model successfully reproduces the differences among real networks.
These results suggest that the presence of societal factors undermines the function of the scientific reward system.
To improve the status quo, we advocate for reforming the reference list format in papers, urging journals to require authors to separately disclose logical references and social references.

\clearpage


\newrefsection

\section*{Main}
Scientific literature system has always been the most central human knowledge system.
It contains a large amount of knowledge, and the citations within it reflect, to some extent, the relationships between knowledge.
Although scientific knowledge is objective, citing knowledge in academic writing has always been considered as a human behaviour, leading to knowledge-oriented citations being mixed with many societal factors\supercite{tahamtanFactorsAffectingNumber2016,mammolaMeasuringInfluenceNonscientific2022}.
Journal reputation\supercite{traagInferringCausalEffect2021}, author reputation\supercite{petersenReputationImpactAcademic2014a}, country\supercite{pasterkampCitationFrequencyBiased2007,gomezLeadingCountriesGlobal2022}, gender\supercite{lariviereBibliometricsGlobalGender2013,teichCitationInequityGendered2022}, and even the length of the paper's title may affect citation behavior\supercite{letchfordAdvantageShortPaper2015};
irresponsible citations abound---more than a quarter of citations are found to be perfunctory or nonessential\supercite{krampenValidityCitationCounting2007,bornmannWhatCitationCounts2008,teplitskiyHowStatusResearch2022};
citation manipulation\supercite{fongAuthorshipCitationManipulation2017,joshiDeceptionManipulatedCitations2024}, such as excessive self-citation\supercite{aksnesMacroStudySelfcitation2003}, reciprocal citation\supercite{fisterDiscoveryCitationCartels2016,zagglManipulationExplicitReputation2017a}, and coercive citation\supercite{wilhiteCoerciveCitationAcademic2012}, occurs frequently, which may eventually slide into the abyss of academic misconduct.
These distorted citation practices have many consequences\supercite{horbachMetaResearchHowProblematic2021}.
Most directly, they interfere with identifying essential connections between knowledge and may contribute to spurious scientific impact;
more broadly, all work that uses citation data for subsequent research is inevitably affected by them.

On this mixed scientific literature system, many studies have been done.
Citing behaviour may derive from the authors' complex motivations---perhaps as many as a dozen---and take on different functions in academic writing\supercite{tahamtanCoreElementsProcess2018,tahamtanWhatCitationCounts2019}.
One succinct division is distinguishing between substantive and rhetorical citing\supercite{teplitskiyHowStatusResearch2022}.
Substantive citing involves theoretical bases, results, methods, etc. that have a meaningful influence on the authors' work;
rhetorical citing, on the contrary, is for purposes such as persuading the reader or enhancing the integrity of content.
On the one hand, related research tries to identify different types of citations based on real data\supercite{zhuMeasuringAcademicInfluence2015,prideIncidentalInfluentialDecade2017,hoppePredictingSubstantiveBiomedical2023}.
On the other hand, the academic consequences of rhetorical citing are observed by pure model simulations\supercite{baoSimulationbasedAnalysisImpact2024a}.
These are both indirect and difficult to verify extensively for a simple reason:
In the real world, there was no scientific literature system that is unadulterated by societal factors for people to observe.

Is there any other ``ideal'' knowledge system, with little or no influence from human behaviour?
We believe that the knowledge system constituted by mathematical theorems is just such a system\supercite{shapiroPhilosophyMathematicsStructure1997}.
Similar to a scientific citation network, a network can be constructed between theorems.
The nodes in the network are theorems or axioms.
If the proof of a theorem uses another previous theorem or axiom, there is a directed link between the two nodes.
Since the relationship between theorems is a strict premise-conclusion relationship guaranteed by deductive reasoning, the links in such a network are purely logic-based\supercite{detlefsenProofItsNature2008}.
Accordingly, similar to the substantive-rhetorical dichotomy, we can reclassify citations into logical and societal citations.
Under this doctrine, theorem networks contain only logical citations, while scientific citation networks have both logical citations based on scientific objectivity and societal citations formed by societal factors.
Therefore, comparing the theorem network and the scientific citation networks enables us to observe how societal factors affect knowledge systems.

\section*{Mathematical theorem network}

As far as we know, it was Stephen Wolfram who first modeled the relationships between mathematical theorems as complex networks.
In his well-known book \textit{A New Kind of Science}, he showed in the form of a complex network how Euclid's \textit{Elements} derived all the subsequent 465 theorems from the five axioms\supercite{wolframNewKindScience2002}.
In recent years, with the development of computer-assisted proof, a series of projects called proof assistant or interactive theorem prover have been developed, such as Lean\supercite{mouraLeanTheoremProver2021}, Coq\supercite{bertotInteractiveTheoremProving2004}, HOL\supercite{harrisonHOLLightOverview2009}, and Isabelle\supercite{paulsonIsabelleGenericTheorem1994}.
Both the proof of the four-colour theorem in the last century\supercite{appelEveryPlanarMap1989} and the recent integration with large language models (LLMs) to accelerate the discovery of mathematical knowledge\supercite{yangLeanDojoTheoremProving2023,songLargeLanguageModels2024} show the great potential of such tools.
Many users are active in the communities of these projects, contributing formal expressions and formal proof processes of many theorems to facilitate others to invoke them for further proofs of more complex theorems.
As a result, these projects have libraries of mathematical theorems
\footnote{These projects have also prompted Wolfram to update his research on theorem networks, see his article ``The Empirical Metamathematics of Euclid and Beyond''\supercite{wolframEmpiricalMetamathematicsEuclid2021}.}.

\begin{figure}[h]
  \centering
  \includegraphics[width=\textwidth]{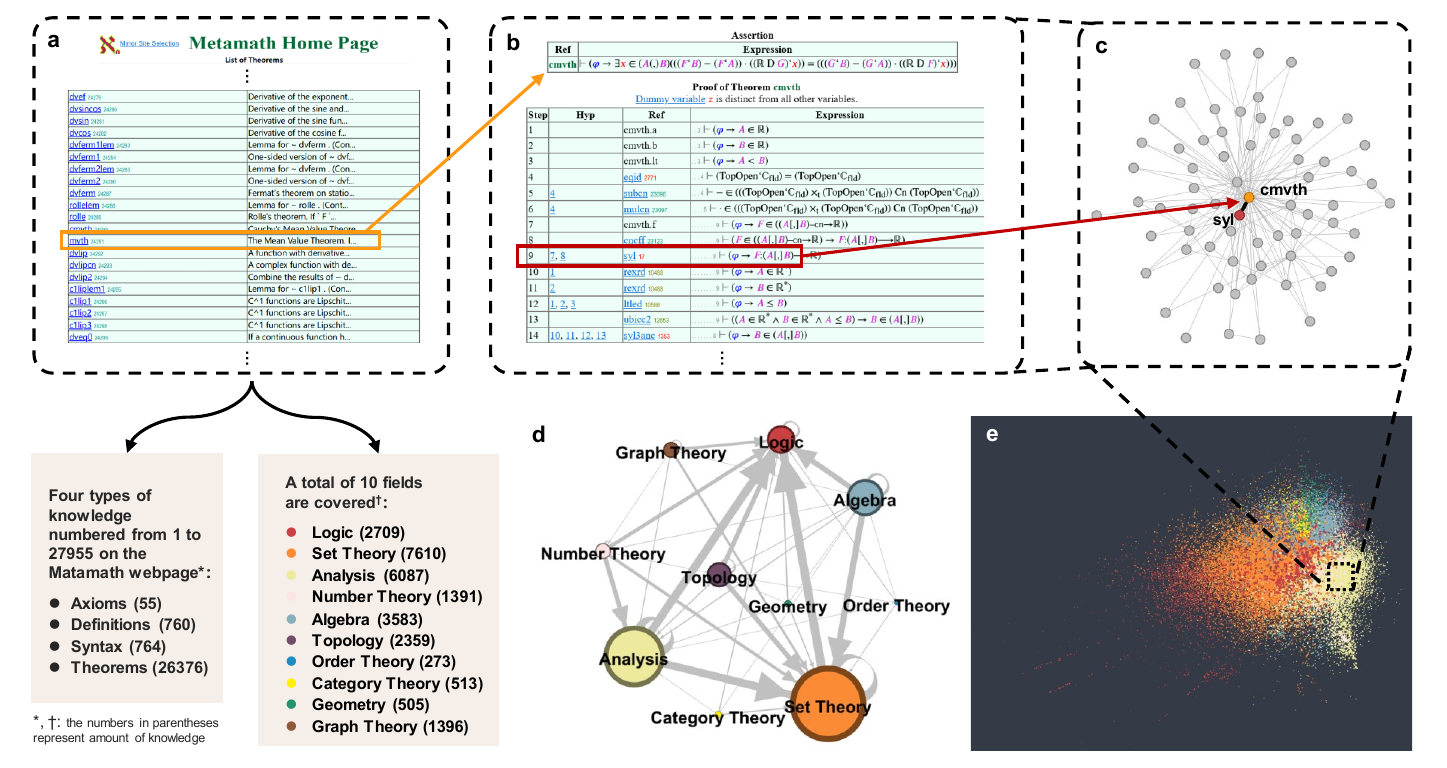}
\caption{\textbf{Scraping citation relationships between mathematical theorems from Metamath web pages.}
  {
  \textbf{a} The Metamath website features a list of mathematical knowledge, including 55 axioms, 760 definitions, 764 syntaxes, and 26,376 theorems.
  We would only consider theorems and axioms in scraping citation relationships.
  \textbf{b} Each entry comes with a dedicated page detailing the formal expression of the knowledge and its proof process (the latter only for theorems).
  This subplot shows part of the webpage for Cauchy's Median Theorem (coded as ``cmvth'').
  \textbf{c} The proof process of Cauchy's Median Theorem is transformed into a network.
  The ninth step uses the theorem coded as ``syl'', thus an edge is created between the two nodes ``cmvth'' and ``syl'' in the network.
  The proof is divided into 158 steps and involves 67 theorems.
  This subplot demonstrates a subgraph induced by these nodes.
  \textbf{d} The whole data covers a total of ten mathematical fields: logic, set theory, analysis, number theory, algebra, topology, order theory, category theory, geometry, and graph theory.
  This subplot shows the interdependence for these ten fields, generated using Gephi.
  The node size is proportional to the number of theorems or axioms in the field, and the edge width is proportional to the number of citations.
  \textbf{e} Visualization of the large-scale network of all 26,426 nodes, generated with Cosmograph.
  Five theorems are excluded from the data as they are deprecated in the Metamath project (although their numbers were retained) and constitute isolated nodes in the network.
  }
  }
\label{figure1}
\end{figure}

In this paper, we extract a mathematical theorem network from Metamath\supercite{megillMetamathComputerLanguage2019}---a same type project for formalization of mathematics--- because it is relatively easy to access.
Following the spirit of Russell and Whitehead's \textit{Principia Mathematica}, Metamath builds a mathematical edifice starting from ZFC set theory and first-order predicate logic, developing the basic knowledge of ten fields, with a total of 55 axioms, 760 definitions, 764 syntaxes, and 26,371 theorems (as of October 2023).
As shown in Fig. \ref{figure1}, we extract and organise them into citation relationships data to construct a theorem network (the number of nodes $N=26,426$, the number of links $M=466,480$).

We mainly compare the theorem network with the mathematical citation network (extracted from SciSciNet\supercite{linSciSciNetLargescaleOpen2023a}, $N=39,028$, $M=169,472$) and the high-energy physics theory citation network\supercite{gehrkeOverview2003KDD2003} (cit-HepTh, $N=27,400$, $M=351,884$) because their network size is close to that of the theorem network, thus avoiding the influence of the network size on the network characteristics.
Both citation networks represent knowledge systems that are mixed with societal factors, and at the same time, the mathematical citation network is closer to mathematics in its various characteristics within the constraints of the discipline.
If our logical-societal dichotomy is reasonable, we expect the nature of the mathematical theorem network to be somewhere between the other two networks.
In addition, we verify the consistency of the results on a large dataset, the APS citation network.
The main text will show the results of the comparison of the theorem network with the mathematical citation network and the cit-HepTh network.
Other results are given in the Extended Data Fig. 7.
See Methods for a detailed description of data preprocessing.

\section*{Differences on Structure}

We begin by examining the most fundamental aspect of these three networks, i.e., structural properties, and identify three key differences: 
out-degree distribution, self-degree correlation, and clustering (comparisons of other structural properties are given in Extended Data Tab. 2).

\begin{figure}[h]
  \centering
  \includegraphics[width=\textwidth]{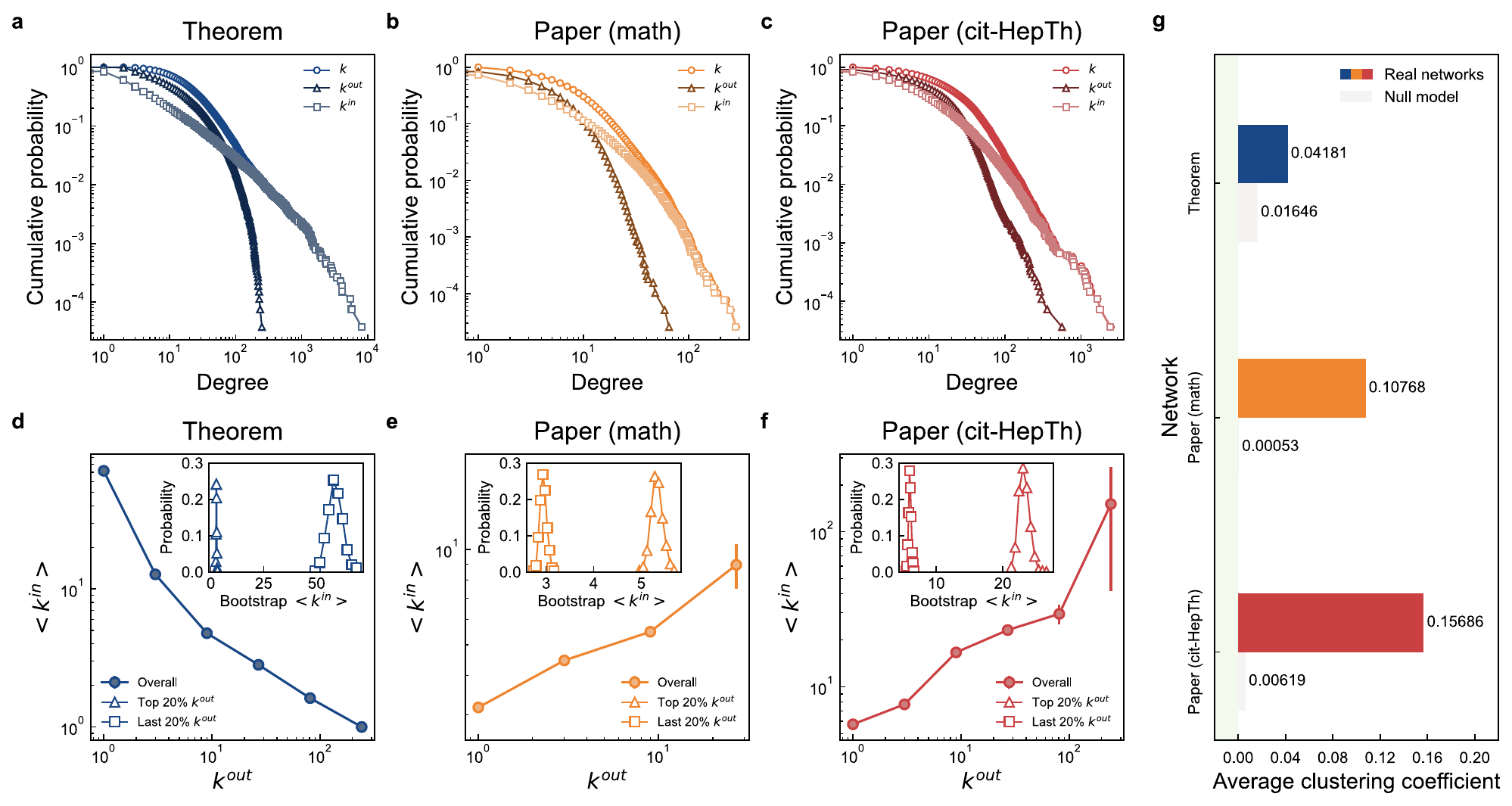}
\caption
{\textbf{Structural differences between the three networks.}
{\textbf{a--c} Cumulative probability distribution of degree of the three networks, by total degree ($k$), out-degree ($k^{out}$), and in-degree ($k^{in}$).
These degree distributions have power-law tails, except for the out-degree distribution of the theorem network, which is closer to an exponential distribution.
\textbf{d--f} Self-degree correlation of the three networks.
The nodes are first binned according to their out-degree, with bin boundaries set at $[3^0, 3^1, 3^2, 3^3, 3^4, 3^5]$ (thus equally spaced in logarithmic coordinates).
The average in-degree ($ \langle k^{in} \rangle$) of each group is then computed and plotted on a line graph.
For instance, the coordinates of the blue dot in the upper leftmost corner of \textbf{d} are $(1, 71.818)$, indicating that the average in-degree for nodes whose out-degree is in the interval $[3^0, 3^1)$ is 71.818.
The error bars represent the standard errors.
It can be seen that the theorem network and the two citation networks have opposite trends.
To further verify this difference, we extracted nodes in the top 20\% and last 20\% of the out-degree and compared the average in-degree of the two groups across the three networks using the bootstrap method.
The insets show the probability distribution of the average in-degree for 1,000 realizations.
It can be seen that in the theorem network, nodes in the top 20\% by out-degree have a lower average in-degree than those in the last 20\%, whereas the opposite is true for the two citation networks.
\textbf{g} The average clustering coefficients for the three networks and their corresponding null models.
The null model was constructed as outlined in the Methods.
The shown values for the null model are the average of the ten null models.
}
}
\label{figure2}
\end{figure}

Firstly, Fig. \ref{figure2}a--c show the total degree ($k$) distribution, in-degree ($k^{in}$) distribution, and out-degree ($k^{out}$) distribution of the three networks.
These networks consistently exhibit power-law tails in their total degree and in-degree distributions.
However, the out-degree distribution of the theorem network is closer to an exponential distribution, whereas that of the two citation networks still has power-law tails.

Secondly, Fig. \ref{figure2}d--f show the self-degree correlation\supercite{anAnalysisUSPatient2018} of nodes across the three networks, which refers to the correlation between the nodes' own out-degree and in-degree.
In the theorem network, a higher out-degree of a node is associated with a lower average in-degree ($ \langle k^{in} \rangle$), while the two citation networks display the opposite trend.
The Spearman's correlation coefficients between out-degree and in-degree for the three networks were -0.325, 0.148, and 0.353, respectively, with all $p$-values less than 0.001.
For each network, we take two groups of nodes---the top 20\% and the last 20\% of nodes in terms of out-degree--- and bootstrap sample them (1,000 realizations) to compare the probability distributions of their average in-degree, as shown in three insets.
It is evident that the average in-degree of nodes with the top 20\% out-degree is lower than that of nodes with the last 20\% out-degree in the theorem network; while the opposite holds true for the two citation networks.
All results of three pairs of bootstrap t-tests\supercite{efronIntroductionBootstrap1993} have $p<0.001$, indicating significant differences in the average in-degree of the two groups of nodes across all networks.
These results suggest that papers listing more references can receive more citations, while mathematical theorems with more complex proof processes are not used more often to prove other theorems.

Thirdly, Fig. \ref{figure2}g presents the average clustering coefficient (in terms of directed networks\supercite{fagioloClusteringComplexDirected2007}) for the three networks and their corresponding null models (the generation of the null model is described in Methods).
The value of the theorem network is relatively low, only 0.042, while the value of the cit-HepTh network reaches as high as 0.157.
The value of the mathematics citation network is between the two, at 0.108, which is also relatively high.
Clustering is often thought to be related to human behaviour, and high clustering occurs in social networks such as acquaintance networks\supercite{davisSmallWorldAmerican2003}, e-mail networks\supercite{ebelScalefreeTopologyEmail2002}, and scientific collaboration networks\supercite{newmanScientificCollaborationNetworks2001}, where the implication is that ``a friend of your friend is also your friend''.
This is echoed in our results: Citation networks mixed with societal factors show relatively high clustering, while the more objective theorem network exhibits relatively low clustering.
As we will see next, these structural differences will inevitably have an impact on the function of the networks.

\section*{Differences on function}

In classical theories of the sociology of science, the institution of science is viewed as a reward system\supercite{mertonSociologyScienceTheoretical1974a}.
According to Merton, when the institution of science works efficiently, recognition and rewards are accrued to those who have made genuinely original contributions, motivating scientists to pursue scientific output\supercite{mertonPrioritiesScientificDiscovery1957}.
This is in line with one's intuition---ideally, it should be the normal function of the institution of science to ensure more original results receive greater recognition.
Since citations are considered an important form of recognition\supercite{coleScientificOutputRecognition1967a}, there should be a positive relationship between a paper's originality and its citation counts in the scientific literature system.

\begin{figure}[h]
  \centering
  \includegraphics[width=\textwidth]{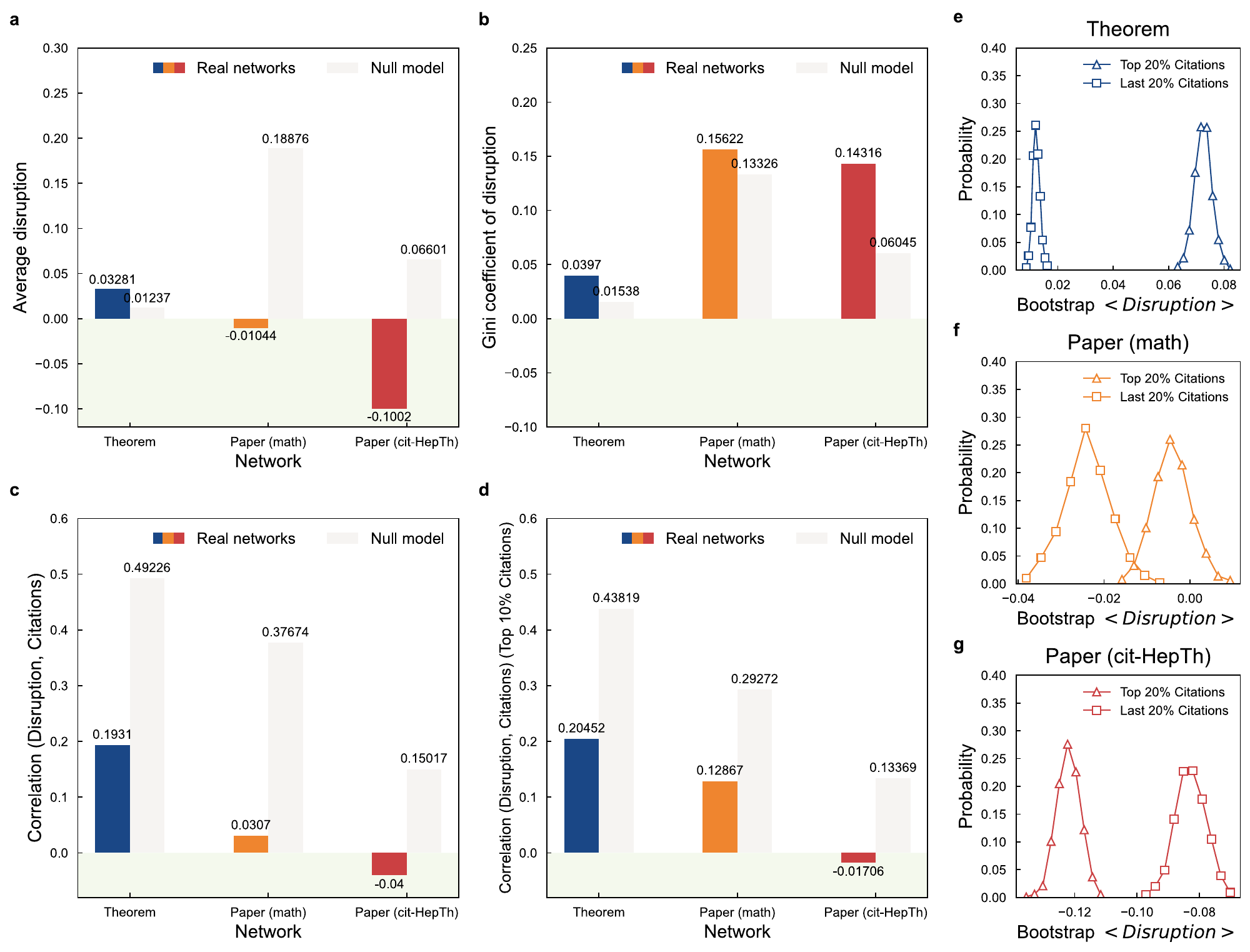}
\caption{\textbf{Functional differences between the three networks.}
  {\textbf{a} The average disruption of the three networks and their corresponding null models.
  They are close to zero in the first two real networks.
  The distribution of disruption for the three networks can be seen in Extended Data Fig. 2.
  Since papers have a temporal attribute while theorems do not, we use the concept of topological generations instead of time to calculate disruption, see Methods for details.
  \textbf{b} Gini coefficient of disruption for the three networks.
  The disruption was normalised to between 0 and 1, then the Gini coefficient was calculated.
  The Gini coefficient for the citation network is relatively higher, implying that there is more inequality in the distribution of disruption.
  \textbf{c} The disruption-citations correlation of the three networks and their corresponding null models.
  Due to the non-skewed distribution of disruption, we directly use the Pearson correlation coefficient to calculate the correlation between disruption and citations.
  All three networks weaken the correlation which is under random circumstances, but the reduction in the theorem network is relatively small compared to the citation network.
  \textbf{d} The disruption-citations correlation of highly cited theorems or papers.
  In the three networks, the top 10\% of cited nodes were selected to calculate the subversion-citation correlation.
  \textbf{e--g} Using bootstrap methods to observe the difference in average disruption between highly cited and lowly cited nodes.
  In each of the three networks, the top 20\% cited nodes and the last 20\% cited nodes were selected, and then the bootstrap method was used to generate 1,000 replications and compute the average disruption to obtain the distribution of average disruption.
  }
  }
\label{figure3}
\end{figure}

We use an indicator that has garnered significant attention in recent years, disruption\supercite{wuLargeTeamsDevelop2019}, to measure originality and thus examine this relationship.
By comparing the disruption-citations correlation in three networks, we can quantitatively observe the impact of societal factors on the functionality of the knowledge system.
Fig. \ref{figure3}c illustrates the disruption-citations correlation for the three networks and their corresponding null models.
In absolute terms, the disruption-citations correlations for the three networks are 0.193, 0.031, and -0.040 respectively, all having $p$-values less than 0.001 (the correlation is calculated using the Pearson correlation coefficient).
The theorem network exhibits some level of positive correlation, and the mathematical citation network and cit-HepTh network exhibit either no correlation or a smaller level of negative correlation
\footnote{The correlation result for the large dataset APS, see Expanded Data Fig. 7.
It is also supported by other literature\supercite{zengDisruptivePapersScience2023} that the disruption of a paper is either uncorrelated or weakly negatively correlated with its citations.
}.
In relative terms, the Z-Scores\supercite{maslovSpecificityStabilityTopology2002} of the disruption-citations correlations for the three networks (measure the extent to which the real network deviates from the null model) are -12.076, -58.283, and -17.274, indicating that all three real networks tend to weaken the original correlation, but the objective theorem network weakens the least.
Fig. \ref{figure3}e--f show the difference in disruption between highly cited and lowly cited nodes by using the bootstrap method across the three networks.
In the mathematical theorem network, the probability distribution of the average disruption of nodes with the top 20\% citations is to the right of nodes with the last 20\% citations (Fig. \ref{figure3}e), indicating that highly cited theorems are generally more disruptive than low-cited theorems.
The cit-HepTh network shows the opposite trend (Fig. \ref{figure3}f).
That is, the normal function of the reward system is realized to some extent on the theorem network, but not on the citation network which is mixed with societal factors.

Another observable phenomenon is that the existence of societal factors exacerbates the inequality of disruption.
Fig. \ref{figure3}b shows the Gini coefficient calculated after normalising the disruption of the three networks to between 0 and 1.
A larger Gini coefficient means that a smaller proportion of nodes occupy a larger share of disruption and thus more inequality.
It can be seen that the Gini coefficients for each of the three networks are 0.040, 0.156, and 0.143.
Therefore, the mathematical theorem network is more equal in the division of disruption.
On the whole, these demonstrate the destruction of the functioning of knowledge systems by societal factors.
Knowledge systems mixed with societal factors, such as the citation system, deviate from the ideal state envisaged by Merton in the face of interference, whereas more objective knowledge systems, such as the theorem system, align more closely with the Merton ideal.

\section*{Modeling differences}

So far, we have observed various structural and functional differences between the different knowledge systems.
Since the obvious distinction between theorem systems and citation systems is the presence or absence of societal factors mixed in, we infer that it is such societal factors that lead to these differences.
However, drawing such conclusions from observational data alone may require consideration of other potential influences.
To strengthen our argumentation further in a straightforward manner, we consider a mechanism model to simulate the differences.
Our goal is to model objective and non-objective citations in a knowledge system starting with the mechanism of network generation\supercite{betzelGenerativeModelsNetwork2017a}, specifically by designing two different types of links: those generated by scientific objectivity and those generated by societal factors.
The theorem network generation model contains only the former type of links, while the citation network generation model mixes these two types of links.
If these generative models are able to exhibit those differences similar to the real networks, we can consider our inference to be a reasonable explanation.

Given that the relationships between theorems are based on logic, we may refer to the former type of links as ``logical links'' and the latter type of links as ``societal links''.
For logical links, we can model logical associations using vector representations\supercite{leDistributedRepresentationsSentences2014}: each entity (theorem or paper) is randomly assigned a 10-dimensional vector, where each dimension represents a domain (reflecting the reality that theorem data contains 10 mathematical subdomains), and the elements are either 0 or 1.
If an entity is related to a certain domain, then the corresponding dimension of its vector is taken to be 1;
otherwise, it is taken to be 0.
If the cosine similarity between the vectors of two entities is high, we assume that the entities are somehow substantially logically related.
The wiring rules for logical links will require that the scope of the reference be restricted to logically related entities.
For societal links, given that the citation network has a higher clustering coefficient, implying more triangles, we consider using a strategy similar to the vertex copying model\supercite{kleinbergWebGraphMeasurements1999} to generate them.

Combining the above ideas, our network generation model is designed as follows:
At each time step $t$, we add a node to the network.
Subsequently, that node goes through an ever-repeating process A.
Each time process A extends a directed link from that node as a logical link.
Each time process A ends it stops with probability $p$, so process A is executed at least once, and could theoretically be repeated indefinitely.
This would directly lead to the exponential out-degree distribution observed in the theorem network.
The target node selection in each process A is analogous to the local world evolving model\supercite{liLocalworldEvolvingNetwork2003}:
The cosine similarity between the newly added node and all the original nodes in the network can be computed, and then a threshold $w$ is set above which nodes are retained as the set of nodes logical related to that node.
Finally, a node is selected from this set to be connected based on the principle of preference attachment.
Once the ever-repeating process A concludes, it proceeds to the next time step $t+1$.
This is the generative model of the mathematical theorem network.

\begin{figure}[h]
  \centering
  \includegraphics[width=\textwidth]{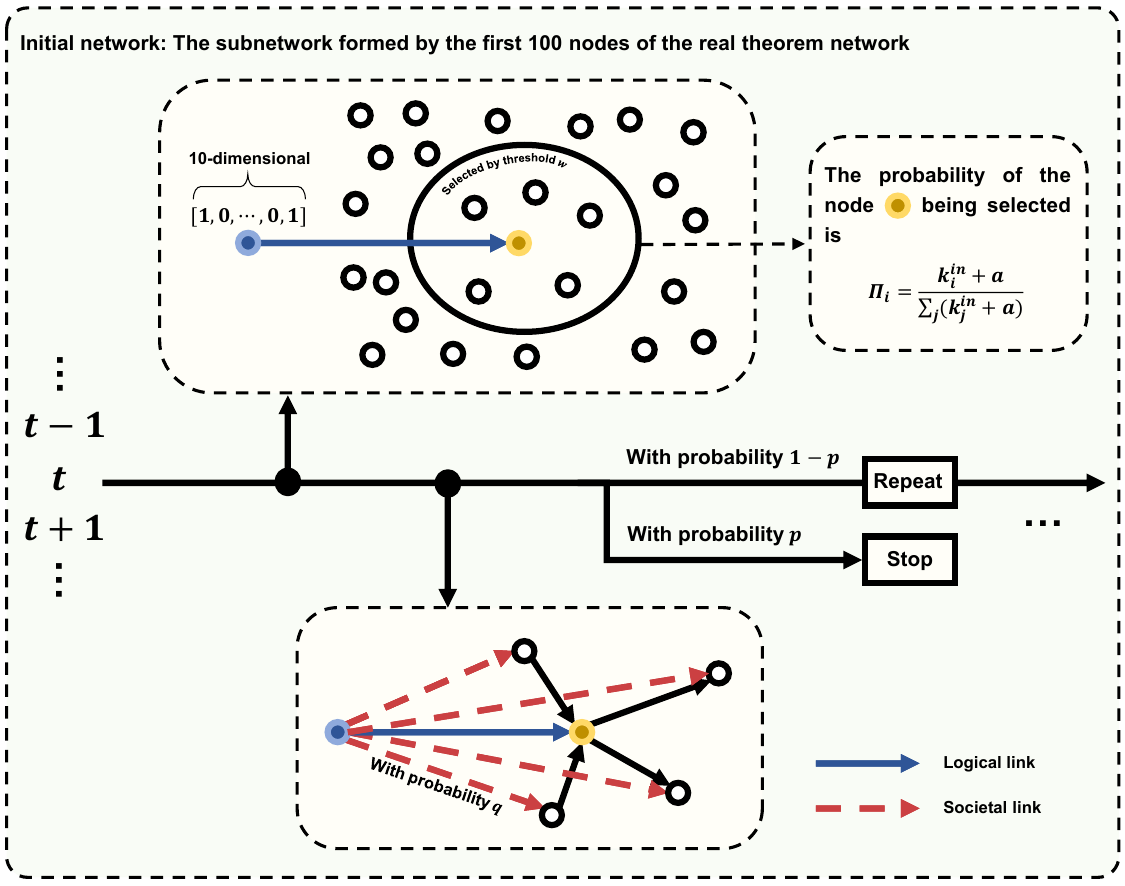}
\caption{\textbf{Design of network generation model.} 
{The initial network is a subnetwork formed by the 100 nodes that first appeared in the real theorem network.
Each time step adds a node to the network, performing the process A shown in the figure.
The four parameters included in the model can be explained as follows:
(1) $p$: Q.E.D. probability.
The repetitive process A stops each time with probability $p$ until it does occur, which can be interpreted as a continuous search for theorems to include in the proof process until the end of the proof.
(2) $w$: logical threshold.
Determines the range of logistic associations.
It affects the number of links $M$ and maximum degree $k_{max}$ of the network.
(3) $a$: initial attractiveness of the nodes.
This arameter in the preference attachment pattern is used to avoid that a node with $k^{in}=0$ can never be connected.
It also affects the $M$ and $k_{max}$ of the network.
(4) $q$: human influence factor.
Affects the number of societal links.
It can measure the degree of influence of societal factors or, in other words, the degree of influence of human behaviour.
}
}
\label{figure4}
\end{figure}

For the generative model of citation networks, an additional part is integrated into each process A:
After selecting the target node of the logical link, all neighbors of that node (both predecessors and successors) are checked to form a neighbor set.
The newly added nodes are then connected to each node in the set with probability $q$, forming an indeterminate number of societal links.
If occasionally, the target node of a logical link is with a large degree, the newly added node will eventually have a large out-degree, resulting in a power-law out-degree distribution like that of a citation network.
Note that when $q=0$, this additional part is effectively null, and the model degenerates into a theorem network generation model.
Thus, we can unify the two cases of theorems and citations using one design scheme, which is illustrated graphically in Fig. \ref{figure4}.
It is a concise model with only four parameters: the Q.E.D. probability $p$, the logical threshold $w$, the initial attractiveness $a$, and the human influence factor $q$.

\begin{figure}[h]
  \centering
  \includegraphics[width=\textwidth]{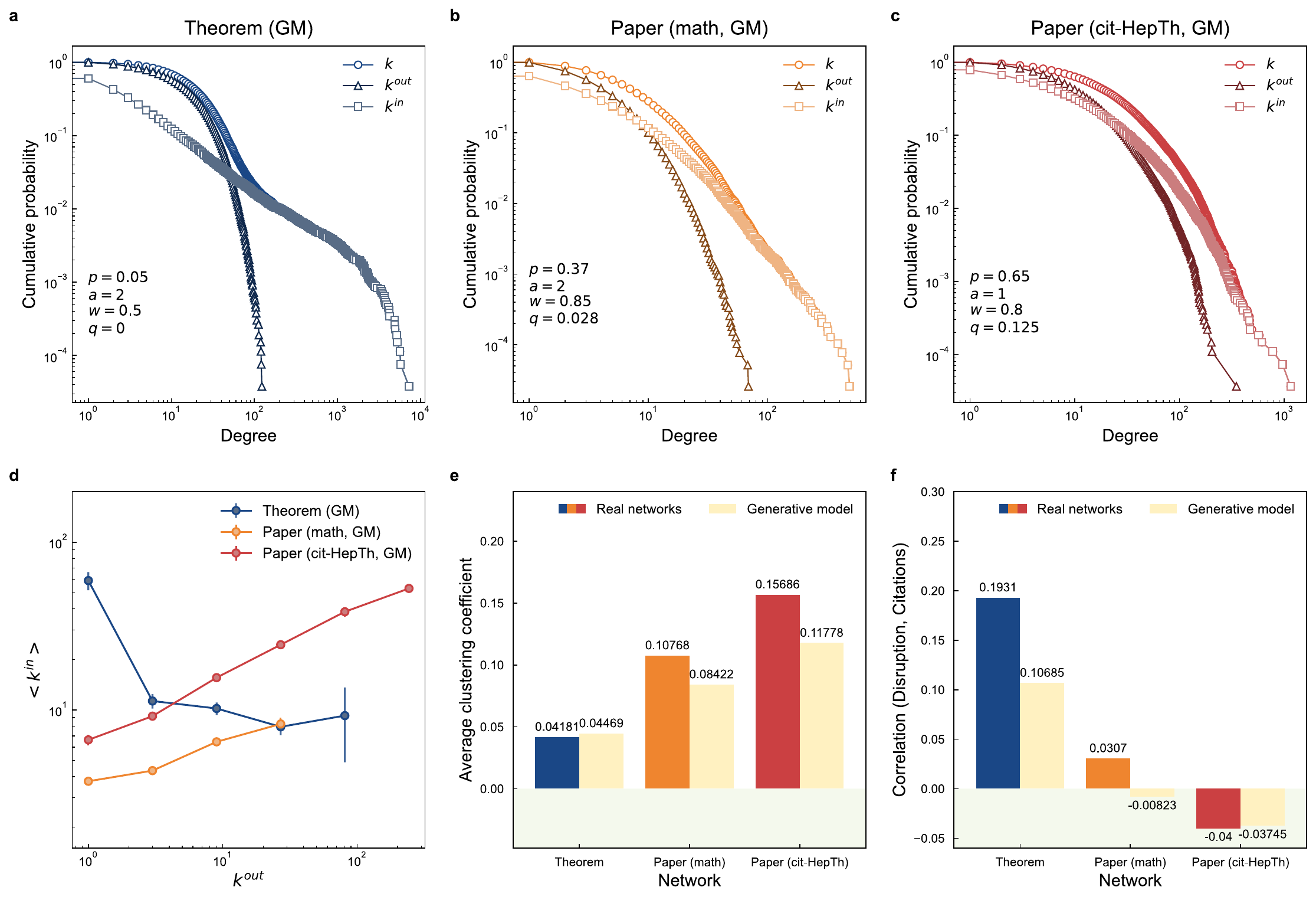}
\caption{\textbf{Reproducing the differences using the network generation model.}
  {The generative model stops when it reaches the same number of nodes as the real network.
  The resulting number of links produced succeeds in being similar to the real network.
  \textbf{a--c} The out-degree distributions of the theorem network generation model and the citation network generation models are significantly different.
  The former is closer to an exponential distribution, while the latter is closer to a power-law distribution.
  \textbf{d} The self-degree correlation of the theorem network generation model is negative, while that of the citation network generation models are positive, consistent with reality.
  The line chart is presented in the same manner as in Fig. \ref{figure2}.
  \textbf{e--f} Gradually increasing clustering coefficients and decreasing disruption-citations correlations from the theorem network generation model, the mathematical citation network generation model, to the cit-HepTh network generation model.
  The values from the generative model remain close to reality.
  }
  }
\label{figure5}
\end{figure}

The model succeeds in reproducing all the differences between the three real networks, as shown in Fig. \ref{figure5}.
The most notable of them is the distinction in the human influence factor $q$, which is 0 for the theorem network generation model ($N=26,426, M=452,641$), 0.028 for the mathematical citation network generation model ($N=39,028, M=171,679$), and 0.125 for the cit-HepTh network generation model ($N=27,400, M=354,259$).
The sequentially increasing $q$ values illustrate the gradual increase in the influence of societal factors from the mathematical theorem system to the citation system.
Without the human influence factor $q$, the generated system is unable to achieve a power-law out-degree distribution, positive self-degree correlation, and high clustering coefficient, which are the characteristics of the citation system.
Yet, it is possible to maintain the proper functioning of the reward function in this case.
Once the human influence factor is mixed in, the features of the citation system emerge, but the reward function is gradually lost (disruption-citations correlation decreases gradually);
and, the more mixed, the more lost.
The level of mixing (the parameter $q$) in the mathematical citation network is in between that of the theorem network and the cit-HepTh network, and thus its nature is also in between.
All these suggest that societal factors play a dominant role in the differences between the three networks and contribute to the anomie of the reward system.

\section*{Discussion}

In this work, we constructed a mathematical theorem network data and compared it with scientific citation networks, discovering their structural and functional differences.
In particular, the citation network exhibits dysfunction, i.e., it fails to allow more disruptive results to receive more recognition, while the theorem network maintains this quality.
We explain this phenomenon by distinguishing between logical and societal citations.
Logical citations are based on scientific objectivity and are present in both theorem network and citation networks;
societal citations are driven by societal factors and exist only in the citation networks, and it is their presence that brings about the observed differences.
This is verified by us using a network generation model.
We thus believe that it is the presence of societal factors causing disruptive papers to fail to get the recognition they deserve.

The success of the modelling mechanism provides us with deeper insights into citation behavior in theorem and citation systems.
Logical citations preferentially connect to highly cited theorems or papers within the logical relevance range, and societal citations further seek out their neighbours for citations.
The latter destroys the purely logical structure between the knowledge represented by the papers and interferes with the citation system's functioning.
In the model, the parameter $q$ determines how many social citations are made. 
From the mathematics theorem network, to the mathematics citation network, to the cit-HepTh network, $q$ increases successively, resulting in greater human influence and more serious functional loss.
Therefore, we expect that $q$ has the potential to be a measure of how much human influence is being exerted.

Another thing worth noting is that, according to the simulation results of the network generation model, most of the papers have only one or two references as their logical premise, while all other references serve other purposes.
For the mathematical citation network generation model, 60.31\% of the papers have only one or two logical citations; for the cit-HepTh network generation model, it is 87.75\%. 
As a whole, most citations in the citation system are societal:
61.14\% of the 171,679 links in the generation model for the mathematical citation network are societal, while 88.25\% of the 354,259 links in the generation model for the cit-HepTh network are societal.
Are we abusing too many societal citations?
We researchers should not be citing casually, perfunctorily, and irresponsibly in our daily academic writing while simultaneously making citation counts to determine the fate of papers and researchers.

Here, we propose a way to improve the status quo.
When authors write their papers, they should disclose to readers which references are logical and which references are societal.
For ease of understanding and dissemination, we might as well refer to logical references as ``core references''.
These are the references that have substantive support on the paper and serve as its foundational premises---those without which the paper cannot stand.
If the scientific community can reach a consensus, journals can reform the reference list format in papers by splitting them into ``core reference list'' and ``other reference list'', thereby requiring authors to disclose them separately.
After a period of time, we could recalculate the various scientometric indicators by considering core and societal references respectively.
Most simply, scholars' citation counts could be split in this way to measure ``logical impact'' and ``social impact''. 
Such an approach might provide deeper insights into the scientific impact of scholars and papers.

\clearpage

\clearpage

\clearpage

\clearpage

\clearpage

\clearpage
\printbibliography

\newrefsection

\clearpage
\section*{Methods}
\subsection*{Data}

\textbf{Mathematical theorem network}. Interactive theorem provers have continuously sparked interest and excitement over the years of their development.
The MetaMath project features contributions from developers and users that include 55 axioms, 760 definitions, 764 syntaxes, and 26,371 theorems (as of October 2023).
These span a total of 10 fields:
set theory, logic, number theory, analysis, algebra, category theory, order theory, topology, geometry, and graph theory.
The specific data collection process has been given in the Results.
A theorem may use the same theorem several times in the process of proving it.
We do not take this into account when constructing the mathematical theorem network, and hence it is a directed unweighted network.
As a formal system, the mathematical theorem network is inherently a directed acyclic graph (DAG).

\textbf{Mathematical citation network}. 
We obtained the names of the top 25 journals in the Mathematics (miscellaneous) category from the renowned journal ranking website SCImago\supercite{gonzalez-pereiraNewApproachMetric2010}.
We excluded eight journals with a favouring of the fields of probability statistics, computer science, and engineering, leaving us with 17 journals.
We believe this list is representative enough, including the most important mathematical journals in the world.
The specific list can be found in the Extended Data Tab. 1.
Next, we screened articles from these 17 journals on SciSciNet\supercite{linSciSciNetLargescaleOpen2023a}, totaling 54,112 articles.
We further extracted the citation relationships between these articles on SciSciNet.

Subsequently, we cleaned the preliminary data as follows:
First, isolated nodes were removed, leaving 39,579 nodes.
Next, the largest connected component was extracted, leaving 39,057 nodes.
Due to data inaccuracies, accidental errors, or very low-probability special cases, the citation network formed by these 39,057 nodes was not a directed acyclic graph.
In order to make it comparable to the mathematical theorem network, we referred to the publication time of papers to remove false links, then detect and remove all remaining cycles in the network.
Finally, the network remained connected and directed acyclic, containing 39,028 nodes and 169,472 links.

\textbf{Cit-HepTh network}.
The cit-HepTh network\supercite{gehrkeOverview2003KDD2003} is a widely used citation network dataset, consisting of articles and their citation relationships in the high energy physics theory category on arXiv from January 1993 to April 2003.
We downloaded it from the Stanford Network Analysis Project (SNAP)\supercite{leskovecSNAPGeneralPurposeNetwork2017}.
The original dataset contains 27,770 nodes and 352,807 links.
After removing isolated nodes, keeping the largest connected component, and removing all cycles, the dataset comprises 27,400 nodes and 351,884 links.

Additionally, in the Extended Data Fig. 7, we analyzed the APS data provided by the American Physical Society\supercite{sinatraQuantifyingEvolutionIndividual2016}.
It comprises 482,566 articles from 1893 to 2010, covering journals including Physical Review Letters, the Physical Review series, and Reviews of Modern Physics.
The APS data underwent the same data-cleaning process, resulting in 468,590 nodes and 5,007,010 links.

\subsection*{Topological sort, topological generations, and null model of DAG}

A DAG has at least one topological sort\supercite{lasserTopologicalOrderingList1961}.
Typically, a topological sort is defined as an ordering of all nodes in a graph such that if there is a directed link from node $i$ (source node) to node $j$ (target node), then node $i$ comes \textbf{before} node $j$ in the ordering, where $i, j \in V$ and $V$ is the set of all nodes in the graph.
Considering that in practical citation networks, the target nodes are the older publications, we adopt an inverse definition here:
A theorem network or a citation network has at least one topological sort such that if there is a directed link from node $i$ to node $j$, then node $i$ comes \textbf{after} node $j$ in the ordering.

Further, we can assign a topological generation to each node, which is a nonnegative integer that measures the layer in which the node is located in the whole network structure, denoted as $g$.
The topological generations must satisfy the condition that if there is a directed link from node $i$ to node $j$, then $g(i) > g(j)$;
and, the topological generation of a node should be its earliest possible generation that they can belong to.
Stated in another way, for all $i \in V$, if we denote the set of successor nodes of node $i$ as $S(i)$, then the topological generation of the node $i$ is
\begin{equation}
g(i) = 
\begin{dcases}
0 & |S(i)|=0\\
max\left\{ g(j), j \in S(i)\right\}+1 & else
\end{dcases}
\label{eq1}
\end{equation}
where $|\cdot|$ is the set cardinality.
For the theorem network, the zeroth generation consists of 55 axioms, as they have no successor nodes.
All other mathematical theorems are logically derived and accumulated from these axioms layer by layer.
All this knowledge is divided into a total of 291 generations, with theorems from different subfields occupying various positions, see Extended Data Fig. 1.
For the mathematical citation network and the cit-HepTh network, there are 6,356 and 2,563 nodes in the zeroth generation, respectively.
They are nodes with $k^{out}=0$.

Assigning topological generations to the nodes in these three networks will help us calculate disruption.
Before presenting this point, please allow us first to introduce the null model of a DAG.
Brian Karrer and Mark Newman\supercite{karrerRandomAcyclicNetworks2009a} constructed a random acyclic graph starting from a topological sort, where each node has its ordering number, specified $k^{out}$, and specified $k^{in}$.
During the random rewiring process, nodes are traversed in the order of the topological sort, and the randomization of outgoing links from each node ensures that they only point to nodes that precede them in the topological sort.
This forms a random acyclic graph.
Here, starting from a specific DAG, we first randomly determine a topological sort, and then proceed with the process devised by Brian Karrer and Mark Newman, thus generating a null model for degree-preserving randomization of the DAG.
Ten null models were generated for each of the three networks.

\subsection*{Disruption}

Traditionally, calculating disruption requires knowing the publication date of a paper\supercite{wuLargeTeamsDevelop2019}.
The theorem network represents an abstract structure where theorems do not have publication dates (one theorem does not correspond to one published paper).
To calculate the disruption of theorems and make them comparable to papers, we consider using the logical order as a substitute for the concept of time, treating the topological generation of a theorem as its ``publication date''.
At the same time, the disruption of a paper is also calculated using its topological generation instead of its publication time.
Specifically, for a focal theorem or paper $i$, we first define an intermediate variable
$\tilde{P}(i) = P(S(i))\cap V_{g>g(i)}$
, where $P(\cdot)$ is the set of predecessor nodes.
Then, the disruption and citations of node $i$ are defined as
\begin{equation}
\begin{dcases}
Disruption(i) = \frac{|P(i)-\tilde{P}(i)|-|P(i)\cap \tilde{P}(i)|}{|P(i)\cup \tilde{P}(i)|}\\
Citations(i) = |P(i)|
\label{eq2}
\end{dcases}
\end{equation}

We can also consider using only the data up to the 10th generation after the appearance of a focal theorem or paper to calculate both.
This approach can help mitigate the bias caused by the cumulative advantage of older theorems or papers to some extent\supercite{sinatraQuantifyingEvolutionIndividual2016, parkPapersPatentsAre2023}.
Specifically, for a focal theorem or paper $i$, we first define intermediate variables
$\grave{P}(i) = P(i)\cap V_{g<g(i)+11}$
and
$\acute{P}(i) = P(S(i))\cap V_{g(i)<g<g(i)+11}$
, then the disruption and citations of node $i$ are defined as
\begin{equation}
\begin{dcases}
Disruption(i) = \frac{|\grave{P}(i)-\acute{P}(i)|-|\grave{P}(i)\cap \acute{P}(i)|}{|\grave{P}(i)\cup \acute{P}(i)|}\\
Citations(i) = |\grave{P}(i)|
\label{eq3}
\end{dcases}
\end{equation}
In the main text, we present the results obtained using Eq. (\ref{eq3}), while the results obtained using Eq. (\ref{eq2}) can be found in the Extended Data Figs. 5 and 6.
The results from both calculation methods are similar, indicating the robustness of the results across different time windows.

To test the validity of replacing time with the topological generations, we examined the differences between the two.
In our data, the mathematical citation data and the APS data contain complete publication date information.
Therefore, we calculated the disruption of this two data using both the publication dates and topological generations.
The results show that the differences between the two methods are minimal, see the Extended Data Fig. 4.

Finally, we explain the calculation of two citation preference metrics\supercite{liProductiveScientistsAre2024a}---reference popularity and reference age.
Reference popularity indicates whether a theorem or paper cites popular or less popular knowledge, measured by the average citation count of its references, i.e.,
Reference age indicates whether a theorem or paper cites newer or older knowledge, measured by the average generational difference between its references and itself.
The formula for these metrics is given by
\begin{equation}
\begin{dcases}
  Reference\ popularity(i) = \frac{1}{|S(i)|}\sum_{j}^{S(i)} |P(j)|\\
  Reference\ age(i) = \frac{1}{|S(i)|}\sum_{j}^{S(i)} g(i)-g(j)
\label{eq4}
\end{dcases}
\end{equation}
Related results are shown in Extended Data Fig. 8.

\section*{Data availability}

The data for the theorem network and the mathematical citation network are available at Github (\url{https:github.com/ffzdshz/theorem_network}).
The cit-HepTh network data can be downloaded from the Stanford Network Analysis Project.
The APS data is not publicly available and requires a request to the American Physical Society for access.
The SciSciNet data can be accessed in Figshare via \url{https://springernature.figshare.com/collections/SciSciNet_A_large-scale_open_data_lake_for_the_science_of_science_research/6076908/1}.

\section*{Code availability}

The analysis in this study was done using Python.
The Python code for analysing networks, calculating metrics, and plotting are available at Github (\url{https:github.com/ffzdshz/theorem_network}).

\printbibliography

\section*{Acknowledgements}

This work was supported by the National Natural Science Foundation of China (Grant No. 72371031).
We would like to thank Qibin Cheng, a PhD student at the School of Mathematics, Renmin University of China, for his contribution to the subfield division of the mathematical theorems;
Huihuang Jiang and Xifeng Gu for their help in data extraction;
Jianlin Zhou, Peng Zhang, and Leyang Xue for helpful suggestions on the manuscript.

\section*{Author contributions}

X.L., A.Z., and Y.F. conceived and designed the study.
X.L. collected the data, processed the data, and performed the analysis.
X.L., A.Z., and Y.F. contributed to the interpretation of the results.
X.L. drafted the manuscript.
A.Z., and Y.F. revised the manuscript.

\section*{Competing interests}

The authors declare no competing interests.

\clearpage
\section*{Extended data}
\captionsetup[figure]{font={stretch=1.2},labelfont={bf},labelformat={empty}}
\captionsetup[table]{font={stretch=1.2},labelfont={bf},labelformat={empty}}

\begin{figure}[h]
  \centering
  \includegraphics[width=\textwidth]{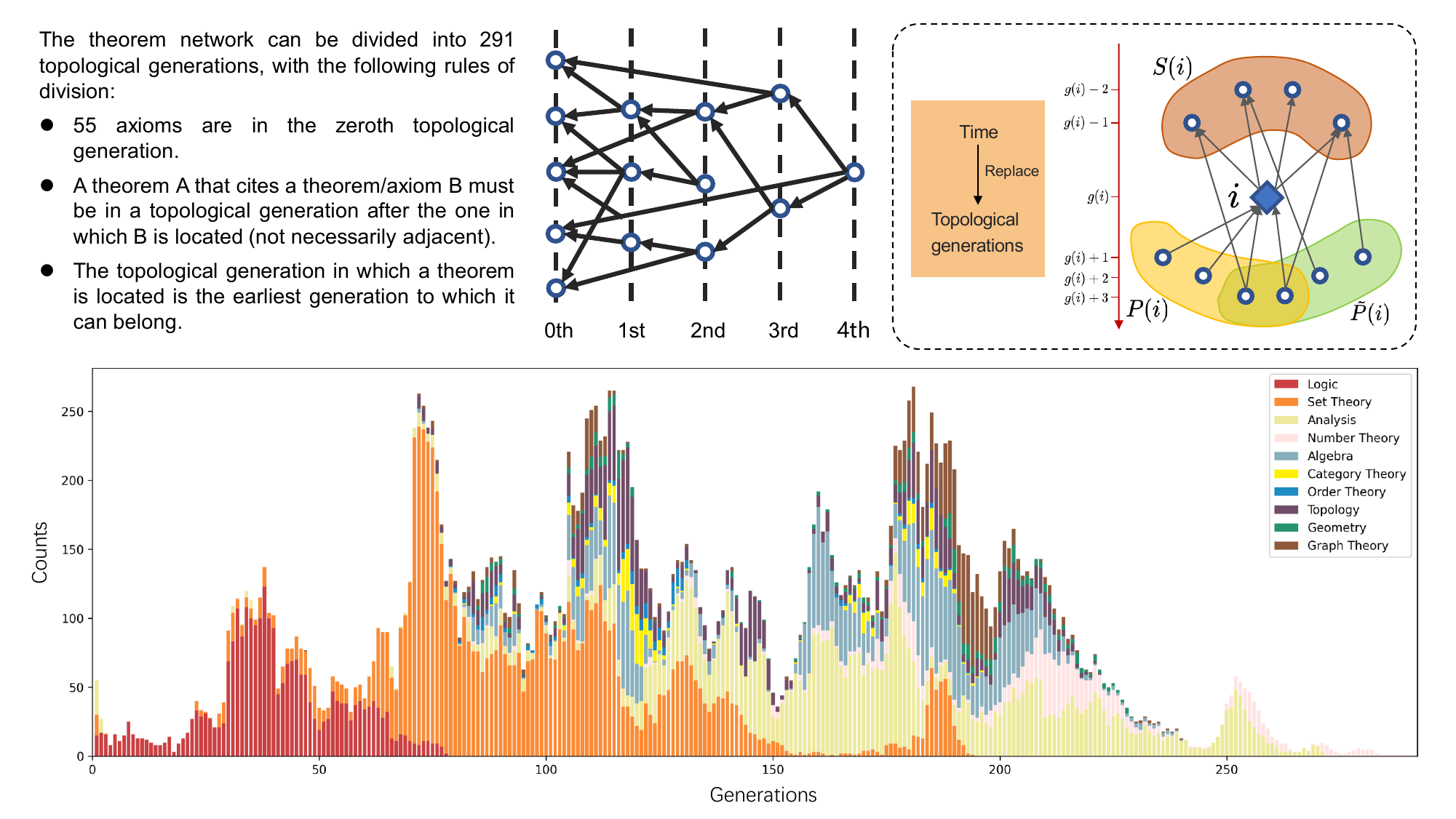}
\caption{\textbf{Extended data Fig. 1 An illustration on topological generations and disruption of theorems.}
  {It should be noted that, whether it is the theorem network or the citation network, outgoing links point from new entities to old entities.
  Therefore, the successor nodes $S(i)$ of node $i$ appear earlier than node $i$.
  }
  }
\label{SMfigure1}
\end{figure}

\clearpage
\begin{figure}[h]
  \centering
  \includegraphics[width=\textwidth]{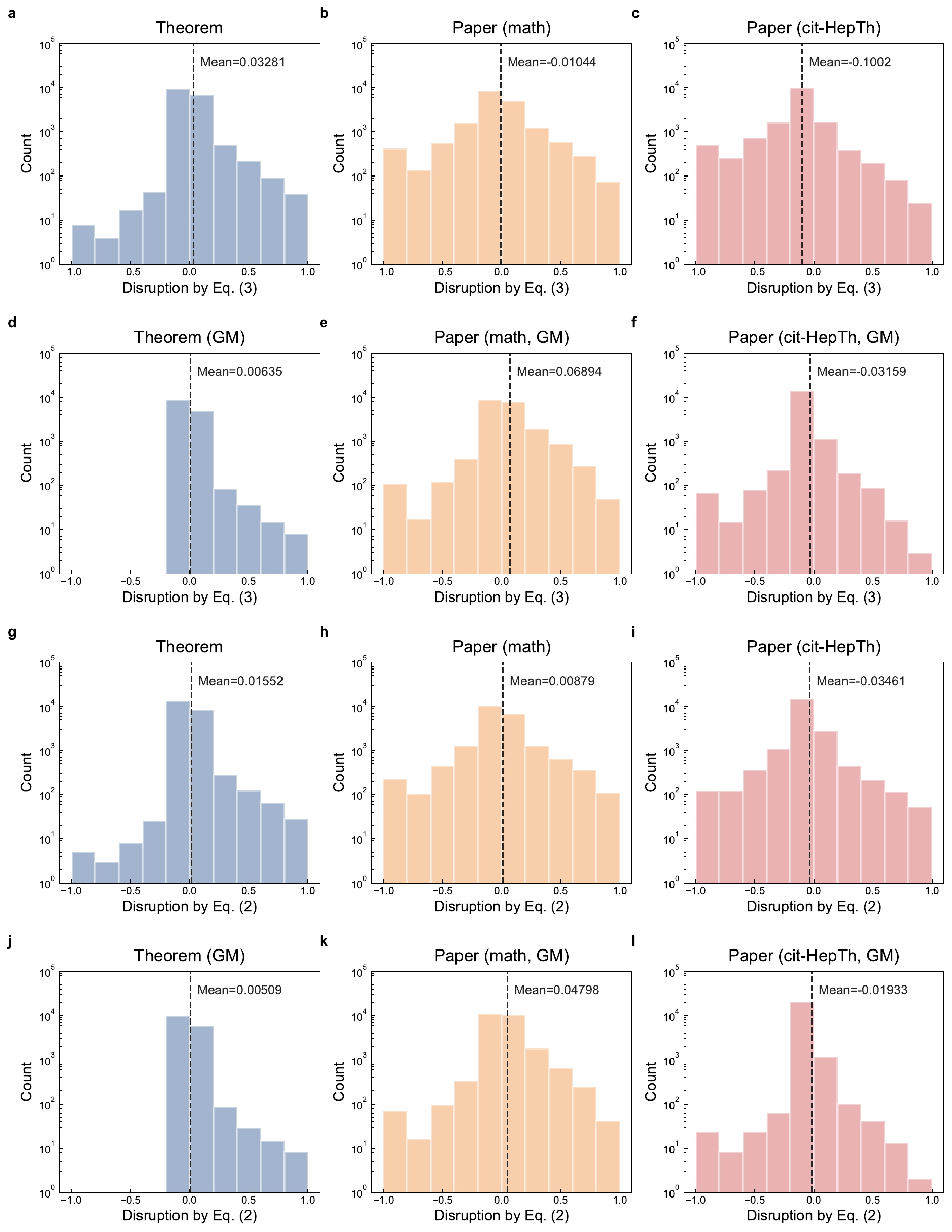}
\caption{\textbf{Extended data Fig. 2 The distribution of disruption.}
  {Subplots \textbf{a}--\textbf{f} represent the six distributions obtained from the disruption computed using Eq. (3), including three real networks and three generative networks.
  Subplots \textbf{g}-\textbf{l} represent the results calculated using Eq. (2).
  }
  }
\label{SMfigure2}
\end{figure}

\clearpage
\begin{figure}[h]
  \centering
  \includegraphics[width=\textwidth]{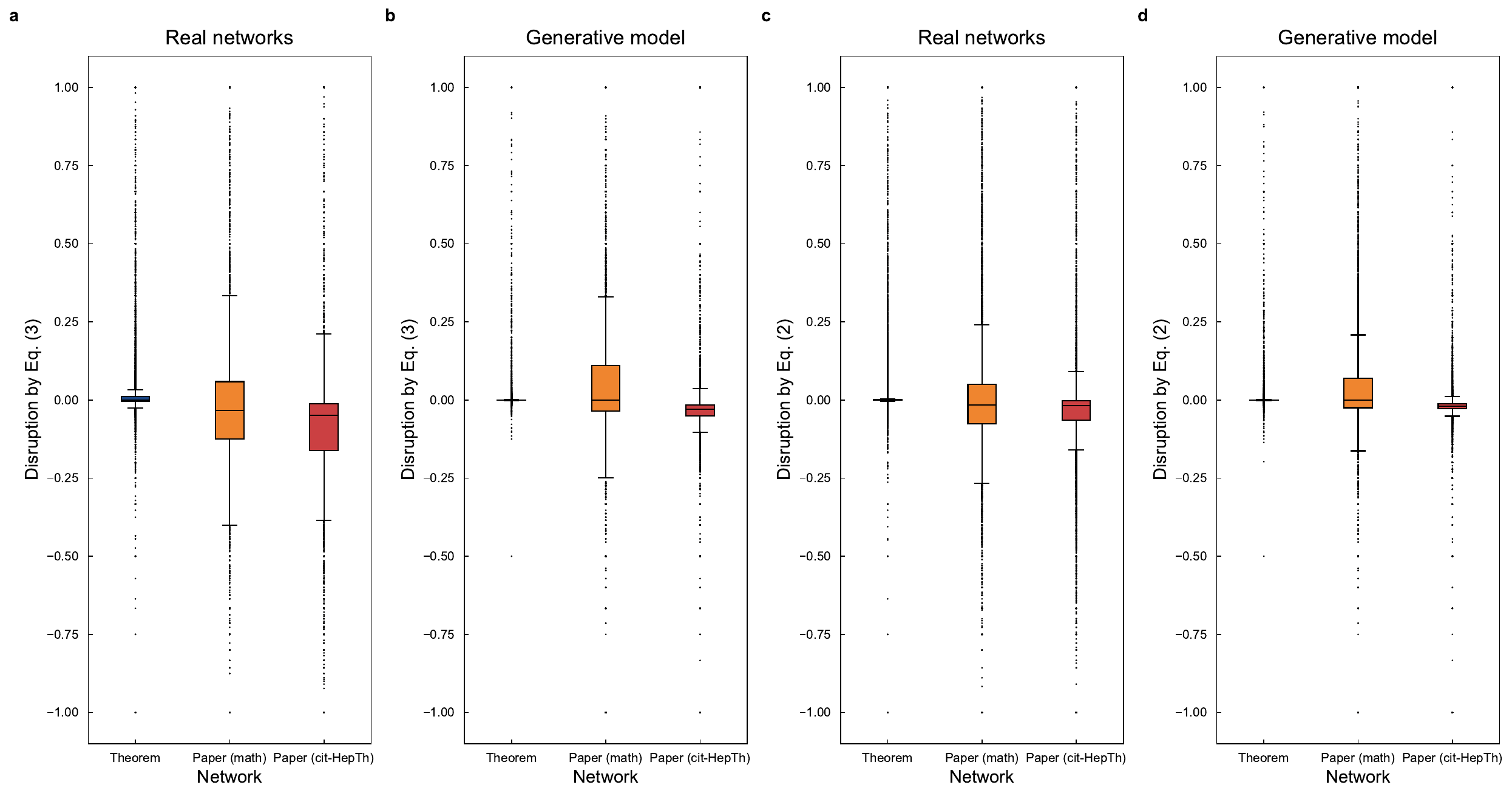}
\caption{\textbf{Extended data Fig. 3 The box plot of disruption.}
  {Subplots \textbf{a} and \textbf{b} represent the two groups of box plots from the disruption of real networks and generative models, computed using Eq. (3).
  Subplots \textbf{c} and \textbf{d} represent the results calculated using Eq. (2).
  }
  }
\label{SMfigure3}
\end{figure}

\clearpage
\begin{figure}[h]
  \centering
  \includegraphics[width=\textwidth]{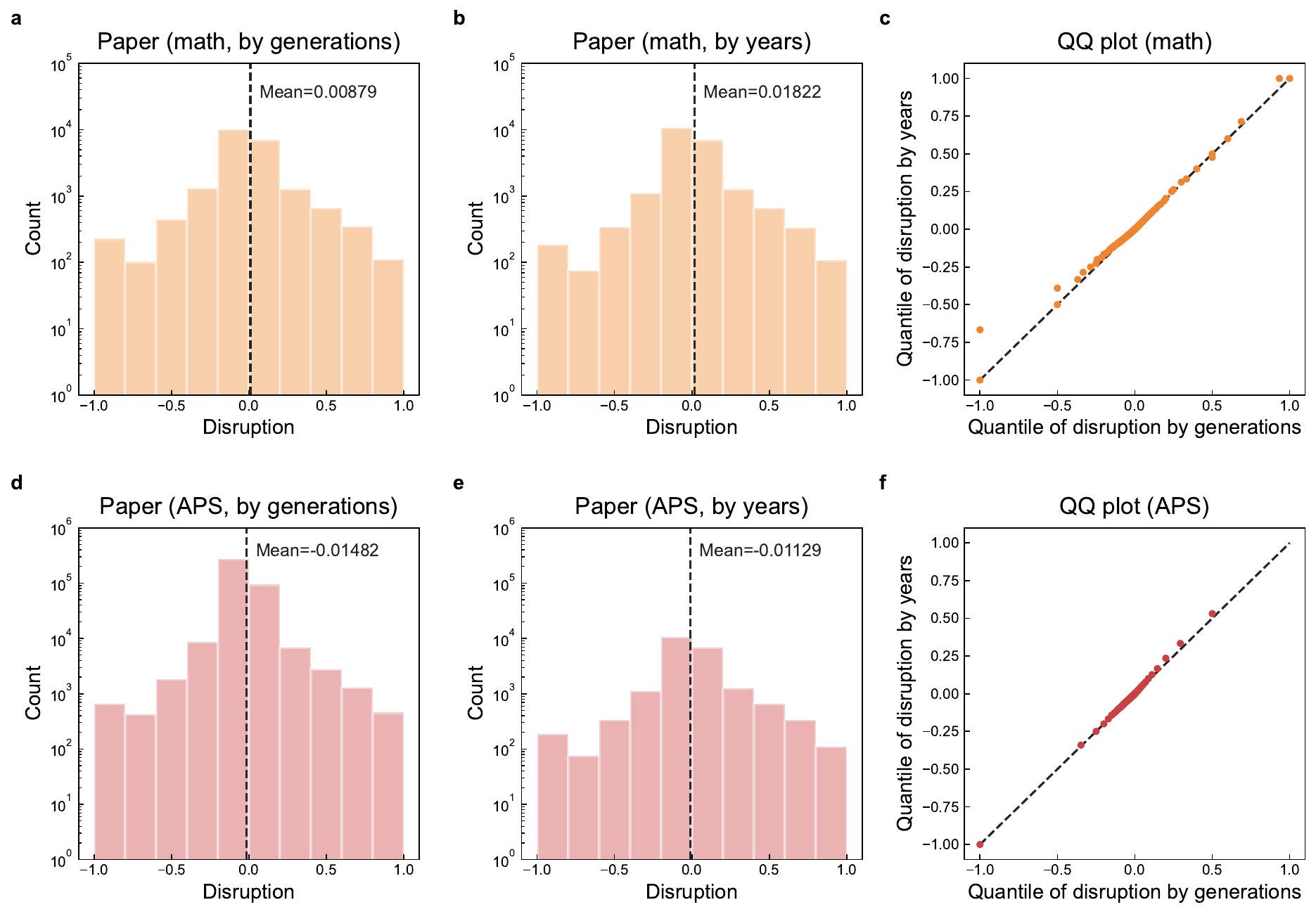}
\caption{\textbf{Extended data Fig. 4 The difference between measuring the disruption of papers through topological generations and time.}
  {Subplots \textbf{a} and \textbf{d} show the distribution of disruption calculated by topological generations for the mathematics citation network and the APS citation network, respectively.
  Subplots \textbf{b} and \textbf{e} display the distribution of disruption calculated by time for the mathematics citation network and the APS citation network, respectively.
  Subplots \textbf{c} and \textbf{f} are Q-Q plots of the distributions obtained from the two calculations to compare whether the two distributions are similar.
  The data lie roughly on a straight line at $y = x$, indicating that the two distributions are similar in shape.
  In addition, the Pearson correlation coefficients, Spearman correlation coefficients, and Kendall rank correlation coefficients between the disruption calculated by topological generations and by time for the mathematical citation network are 0.913, 0.978, and 0.895, respectively.
  For the APS citation network, they are 0.930, 0.989, and 0.931, respectively.
  The above disruptions are calculated using Eq. (2).
  }
  }
\label{SMfigure4}
\end{figure}

\clearpage
\begin{figure}[h]
  \centering
  \includegraphics[width=\textwidth]{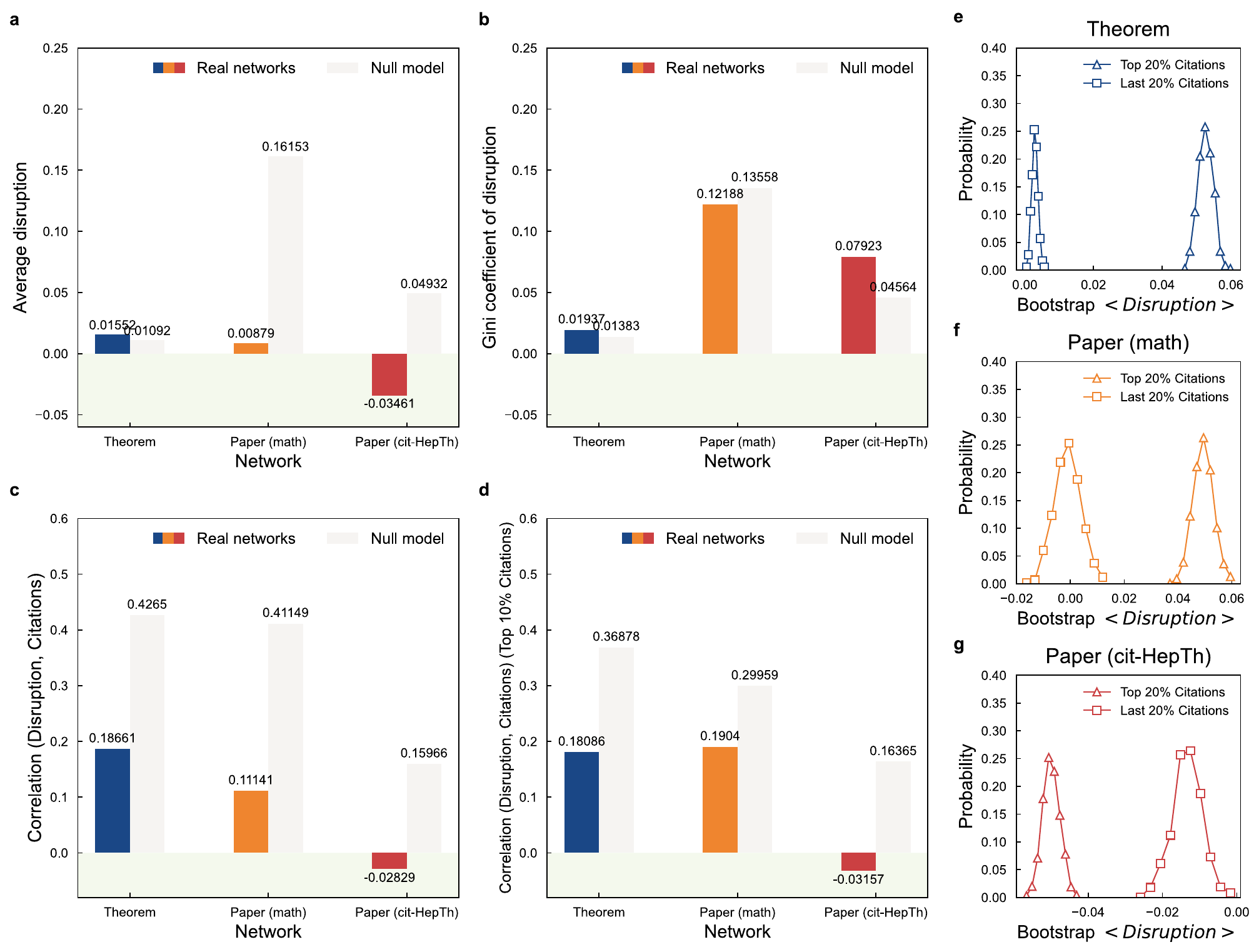}
\caption{\textbf{Extended data Fig. 5 The disruption-citations correlation measured by using Eq. (2).}
  {Compared with Eq. (3), the disruption-citation correlation of the mathematics citation network has increased, but it is still between the theorem network and the cit-HepTh network.
  The Z-Scores of the three networks are -11.769, -53.964, and -22.256.
  }
  }
\label{SMfigure5}
\end{figure}

\clearpage
\begin{figure}[h]
  \centering
  \includegraphics[width=\textwidth]{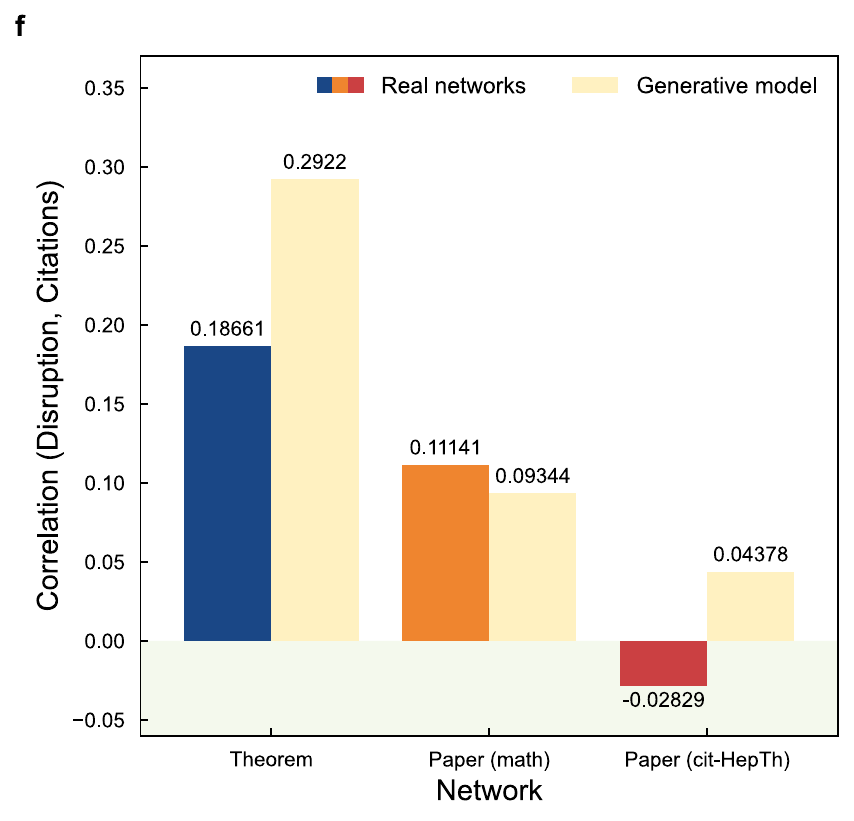}
\caption{\textbf{Extended data Fig. 6 Disruption-citations correlation between the real networks and the generated models computed using Eq. (2).}
  {
  }
  }
\label{SMfigure6}
\end{figure}

\clearpage
\begin{figure}[h]
  \centering
  \includegraphics[width=\textwidth]{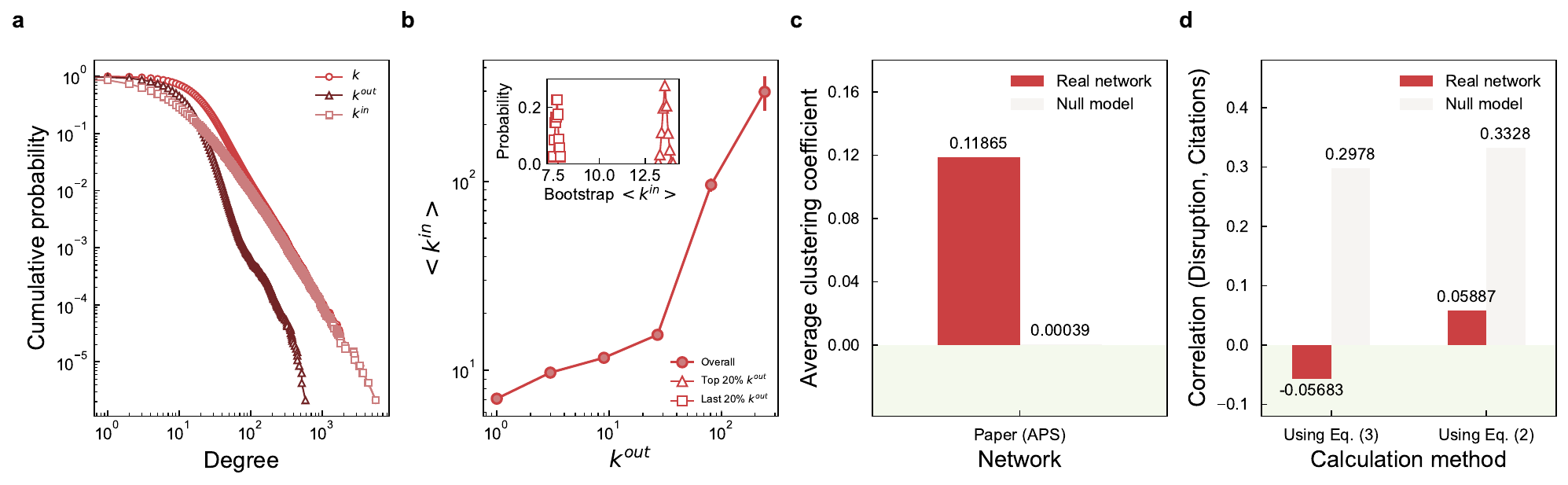}
\caption{\textbf{Extended data Fig. 7 The structure and function of the APS citation network.}
  {Subplots \textbf{a}--\textbf{c} show the three structural features of interest to us in the APS citation system.
  Subplot \textbf{d} examines the functioning of the APS citation system by calculating disruption using Eq. (3) and Eq. (2), respectively.
  These results suggest that the structural and functional features exhibited in the large citation system are similar to the two small citation systems used in our main text.
  }
  }
\label{SMfigure7}
\end{figure}

\clearpage
\begin{figure}[h]
  \centering
  \includegraphics[width=\textwidth]{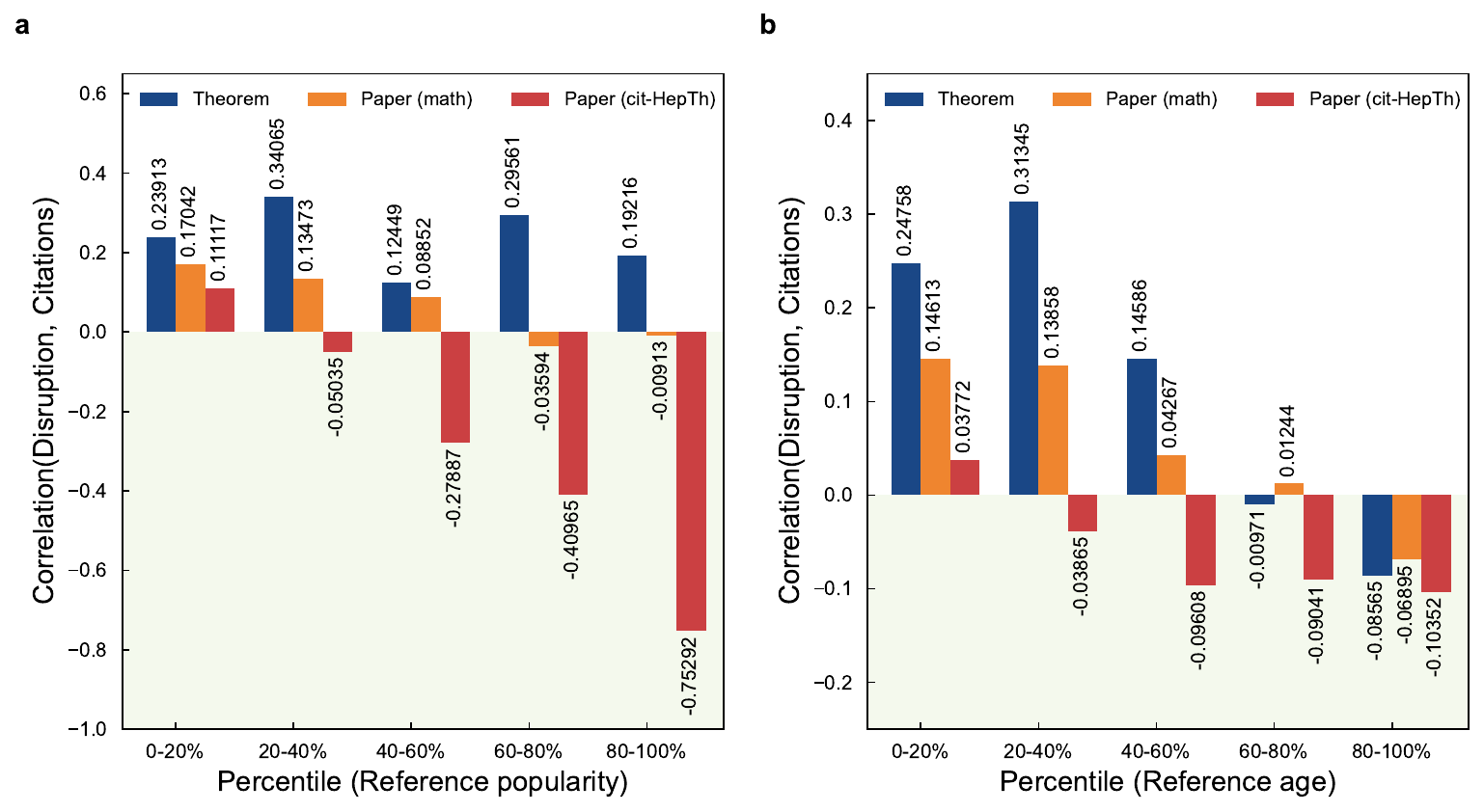}
\caption{\textbf{Extended data Fig. 8 The effect of human citation preferences on citation system function.}
  {Does a theorem or literature prefer to cite more or less popular results (reference popularity); older or newer results (reference age), affect the functionality of the knowledge system?
  Reference popularity can be measured by the average number of citations to a theorem or literature's references;
  reference age can be measured by the average difference in topological generations between a theorem or literature and its references (see Methods).
  Subplots \textbf{a} and \textbf{b} show the disruption-citations correlation at different reference popularity percentiles and different reference age percentiles, respectively.
  In subplots \textbf{a}, nodes are divided into five groups based on the reference popularity, and the disruption-citation correlation is calculated for each group.
  The theorem network exhibits more stability in this aspect, while papers with a preference for citing popular papers show a significant functional decline.
  For the 20\% of papers with the highest reference popularity, the disruption-citations correlation is as low as -0.753.
  This indicates that human citation preferences can interfere with the citation system's functionality.
  However, as can be seen in \textbf{b}, reference age interferes with the function of both the citation system and the theorem system.
  }
  }
\label{SMfigure8}
\end{figure}

\clearpage
\begin{figure}[h]
  \centering
  \includegraphics[width=\textwidth]{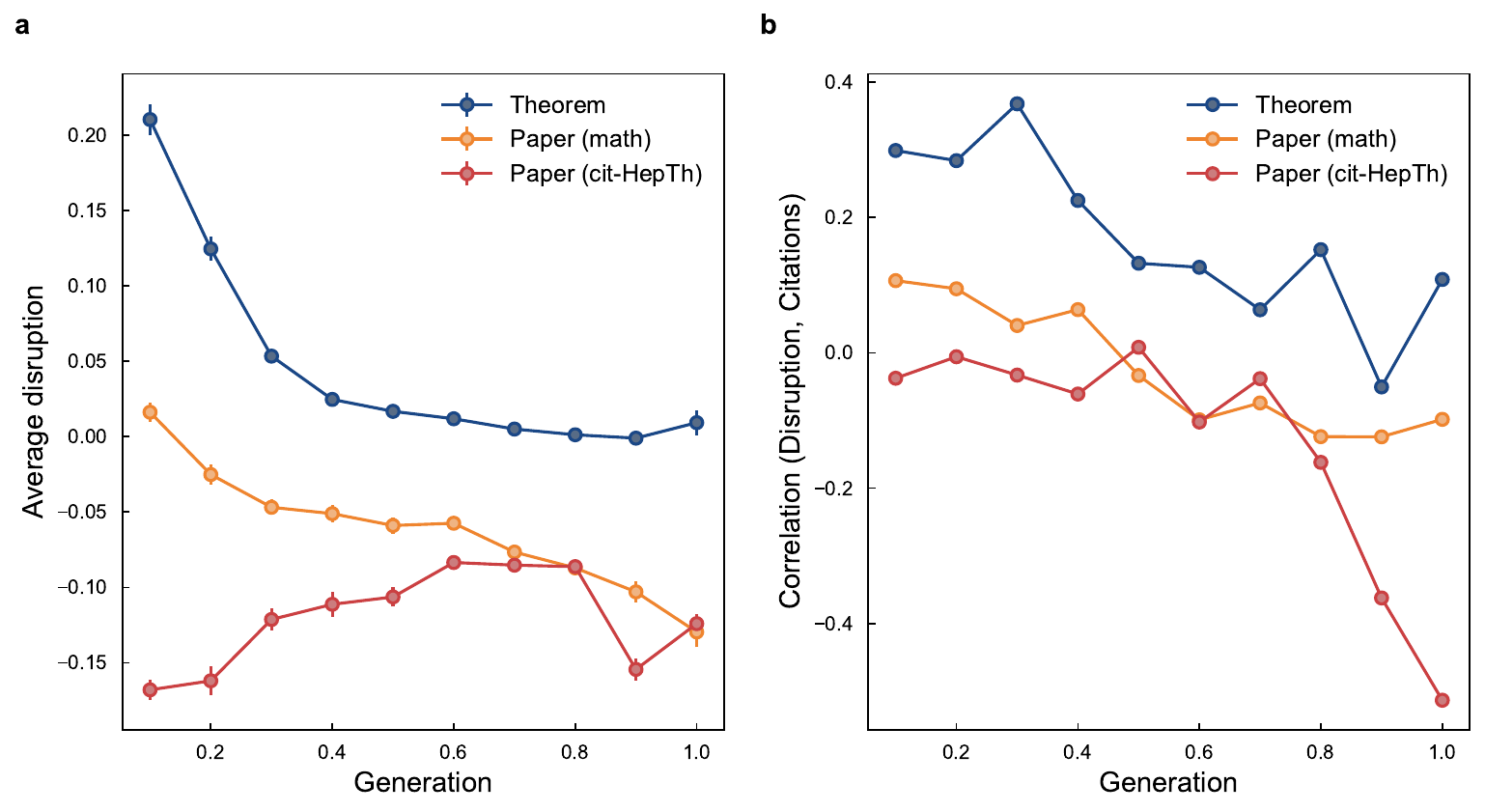}
\caption{\textbf{Extended data Fig. 9  Evolution of disruption and disruption-citations correlation.}
  {After removing the nodes of the first ten generations and the last ten generations, the remaining nodes in each network can be divided into ten groups according to the values of the topological generation.
  For example, the theorem network has a total of 291 generations, and after removing the first ten generations and the last ten generations, there are 271 generations left.
  Therefore, the length of each group is 27.1 generations;
  nodes with topological generations in the interval $[10, 37.1)$ will belong to the first group;
  nodes with topological generations in the interval $[37.1, 64.2)$ will belong to the second group, and so on.
  Subplot \textbf{a} shows the average disruption for each group, while subplot \textbf{b} shows the disruption-citations correlation for each group.
  X-axis scales 0.1, \ldots, 1 represent, in order, from the first group to the tenth group.
  }
  }
\label{SMfigure9}
\end{figure}

\clearpage
\begin{table}[ht]
  \centering
  \caption{\textbf{Extended data Tab.1 Seventeen journals involved in the mathematical citation network.}
  {They comprise a representative range of prestigious and comprehensive mathematical publications.
  The second column gives the number of papers from these journals in the SciSciNet dataset.
  The number of nodes in the cleaned maths citation network is less than the total number of papers shown here.
  }
  }
  \begin{tabular}{@{}cc@{}}
    \toprule
    Journal name & Number of papers \\ 
    \midrule
    Advances in Mathematics & 8,490 \\
    Mathematische Annalen & 8,299 \\
    Duke Mathematical Journal & 6,090 \\
    American Journal of Mathematics & 5,605 \\
    Proceedings of The London Mathematical Society & 5,512 \\
    Annals of Mathematics & 5,298 \\
    Inventiones Mathematicae & 4,145 \\
    Communications on Pure and Applied Mathematics & 2,438 \\
    Journal des Mathematiques Pures et Appliquees & 1,748 \\
    Acta Mathematica & 1,303 \\
    Journal of the European Mathematical Society & 1,215 \\
    Annales Scientifiques De L Ecole Normale Superieure & 1,163 \\
    Journal of the American Mathematical Society & 999 \\
    Advanced Nonlinear Studies & 785 \\
    Publications Mathematiques de l'IHES & 538 \\
    Memoirs of the American Mathematical Society & 343 \\
    Acta Numerica & 141 \\
    \midrule 
    Total & 54,112 \\ 
    \bottomrule
  \end{tabular}
\end{table}

\clearpage
\begin{table}[ht]
  \centering
  \caption{\textbf{Extended data Tab.2 Topological property of the three networks.}}
  \begin{tabular}{@{}cccc@{}}
    \toprule
    Topological property & Theorem & Paper(math) & Paper(cit-HepTh) \\
    \midrule
    Number of nodes & 26,426 & 39,028 & 27,400 \\
    Number of links & 466,480 & 169,472 & 351,884 \\
    Density & $6.68016\times10^{-4}$ & $1.11264\times10^{-4}$ & $4.68721\times10^{-4}$ \\
    Maximum degree & 8,133 & 283 & 2,468 \\
    Maximum out-degree & 250 & 65 & 562 \\
    Maximum in-degree & 8,131 & 279 & 2,414 \\
    Average degree & 35.30462 & 8.68464 & 25.68496 \\
    Degree distribution & power-law & power-law & power-law \\
    Out-degree distribution & exponential & power-law & power-law \\
    In-degree distribution & power-law & power-law & power-law \\
    Average path length & 2.87008 & 6.06854 & 4.27880 \\
    Global clustering coefficient (undirected) & 0.00566 & 0.13648 & 0.11942 \\
    Global clustering coefficient (directed)  & 0.01792 & 0.11229 & 0.13205 \\
    Average clustering coefficient (undirected) & 0.08362 & 0.21537 & 0.31371 \\
    Average clustering coefficient (directed) & 0.04181 & 0.10768 & 0.15687 \\
    Assortativity coefficient (undirected) & -0.06090 & 0.050571 & 0.00135 \\
    Assortativity coefficient (out-out) & 0.01783 & 0.29310 & 0.09526 \\
    Assortativity coefficient (out-in) & -0.06090 & 0.050571 & 0.00135 \\
    Assortativity coefficient (in-out) & -0.01175 & -0.02673 & 0.00546 \\
    Assortativity coefficient (in-in) & 0.01445 & 0.12144 & 0.04062 \\
    \bottomrule
  \end{tabular}
\end{table}

\end{document}